\documentstyle[12pt]{article}

       \def\de{depth}

\let\a=\alpha \let\be=\beta  \let\de=\delta
   
  \let\la=\lambda 
    
\let\om=\omega 
  \let\PH=\Phi

\def\0{\over } \def\1{\vec }     \def\2{{1\over2}} \def\4{{1\over4}}
\def\5{\bar }  \def\6{\partial } \def\7#1{{#1}\llap{/}}
\def\8#1{{\textstyle{#1}}}       \def\9#1{{\bf {#1}}}
 \def\llp{\hbox to 0pt{\hss /\hskip1.5pt}}
\def\llo{\hbox to 0.2pt{\hss /}} \def\llq{\hbox to 0pt{\hss /\hskip0.5pt}}
\def\so{\supset\hbox to 0pt{\hss $\displaystyle -$\hskip1pt}}

\def\<{\langle } \def\>{\rangle }

   \let\hc=\dagger

\def\i{{\rm i}} 

\let\nn=\nonumber
\def\bea{\begin{eqnarray}} \def\eea{\end{eqnarray}}
\def\beann{\begin{eqnarray*}} \def\eeann{\end{eqnarray*}}
\def\beq{\begin{equation}} \def\eeq{\end{equation}}

\date{}
\title{
{\normalsize\rm OUTP 95-21 P}\hfill{\normalsize\rm June 1995}
\vspace*{3.0cm}\\
The Sphaleron Rate in the ``Symmetric" \\
Electroweak Phase}
\author{
O. Philipsen\\
{\normalsize\it Theoretical Physics, University of Oxford,
1 Keble Road, Oxford OX1 3NP, UK}\\
\vspace*{2.0cm}\\
}
\begin{document}
\setlength{\baselineskip}{18pt}
\maketitle
\begin{abstract}
\thispagestyle{empty}
\noindent
I calculate the rate of $B+L$ violating processes in the ``symmetric" phase
of the Standard Model by means of semiclassical methods,
which are based on an
appropriately resummed loop expansion within the dimensionally reduced
theory.
The rate is found to reach its asymptotic form
$\Gamma/V=c(\a_W T)^4$ at a temperature about three times the
critical temperature of the
electroweak phase transition, with a coefficient $c\sim 0.01$
at the one-loop level. The order of magnitude of $c$ is sensitive to
higher order corrections to the dynamically generated
vector boson mass if these are greater than
20\%. The temperature below which baryon number dissipation is in thermal
equilibrium in the early universe is estimated
to be $T_*\sim 10^{11} {\rm GeV}$.
\end{abstract}
\setcounter{page}{0}

\pagestyle{plain}
\newpage
\noindent
The origin of the observed baryon asymmetry of the universe
still remains a puzzle, despite many attempts to explain
it \cite{ckn93}.
Whatever be the mechanism for the generation of the
baryon asymmetry realized by nature, its final value was determined by the
$B+L$ violating processes within the Standard Model which are active
at the epoch of the electroweak phase transition \cite{krs85}.
A particularly important quantity is the
rate at which $B+L$ violation proceeds in the
``symmetric" phase of the Standard Model.
This quantity determines, together with the set of conserved charges,
how much of any primordial $B+L$ and/or B excess survives until the
universe undergoes the electroweak phase transition.  Furthermore, it
also enters all scenarios of electroweak baryogenesis \cite{ckn93}.

As a starting point for discussions of B+L violation at high temperatures
it is customary to approximate the
electroweak theory by an SU(2) Higgs model.
In the broken phase this theory has a
saddle point solution of the static classical field equations, the
sphaleron \cite{km84},
representing the top of a potential barrier
between topologically different vacuum sectors.
Its energy is given by
\beq \label{esp}
E_S=B(\la/g^2)\frac{2m(T)}{\alpha_W}\; ,
\eeq
where $B(\la/g^2)$ is a slowly varying function of the ratio of the
coupling constants and $m(T)$ denotes the
temperature dependent vector boson mass. The rate of B+L violating
processes is related to
the rate at which the thermally excited gauge Higgs system crosses the
barrier, which
may be calculated semiclassically \cite{amcl87}-\cite{dia95} by means
of the Langer-Affleck theory \cite{la69} according to
\beq \label{langer}
\Gamma=\frac{\mid \omega_- \mid}{\pi T}\ {\rm Im} F
\approx \frac{\mid \omega_- \mid}{\pi}\frac{{\rm Im} Z_{S}}{Z_0}\; .
\eeq
Here the free energy $F$ of the system is evaluated by
expanding in Gaussian fluctuations around the sphaleron and the vacuum
solutions, respectively, and $\omega_-^2$ denotes the negative eigenvalue
of the unstable mode.
In the symmetric phase, due to the vanishing of the
vacuum expectation value and the vector boson mass, no
non-trivial solution of the
classical field equations is known, the loop expansion breaks
down and a semiclassical
calculation seems impossible. The general form of the rate may
nevertheless be inferred from dimensional analysis to be \cite{amcl87,ks88}
\beq \label{rate}
\frac{\Gamma}{V}=c\ (\alpha_W T)^4 \ .
\eeq
The dimensionless coefficient $c$ was found in lattice simulations
to be of order ${\cal O}(0.1)-{\cal O}(1)$
\cite{amb90}\footnote{For a more detailed review of the
situation in the symmetric phase see \cite{op93}}.
However, these simulations have been
criticised recently \cite{bod95} and it is not clear how reliable
this result for $c$ is. It is therefore particularly desirable to have
an independent analytical calculation of the rate in the symmetric phase.

A semiclassical calculation for temperatures above the critical
temperature $T_c$ of the electroweak phase transition becomes possible
if a transverse mass $m=m_0 g^2 T$ is generated dynamically for the
spatial components of the vector field.  Such a ``magnetic mass"
screens infrared singularities  of the order $\sim g^2T$,
thus making a loop expansion possible. However, the coefficient
$m_0$ of
this mass is entirely non-perturbative \cite{li80}.
With a massive vector boson, one expects
again a potential barrier
between the different winding number sectors
of the theory, peaked at a saddle point
providing the means to use semiclassical
techniques. On the other hand, the presence of a barrier
leads to a Boltzmann factor $\exp(-\beta E_S)$ at inverse temperatures
$\be=1/T$
suppressing the rate of transitions \cite{krs85}.
With a temperature dependent mass $m\sim g^2T$, the sphaleron energy
is of the form
\beq
E_S= 8\pi m_0 B(\la/g^2)T\; .
\eeq
Consequently, the Boltzmann factor is temperature
independent and the strength of the supression depends
exponentially on the coefficients
of the vector boson mass, $m_0$, and the sphaleron energy, $B$.
Based on this observation a non-linear
sigma model describing massive vector bosons was studied in
Ref.~\cite{op93a}, and the rate
of sphaleron transitions was estimated as a function of the
coefficient of the magnetic mass. The main motivation for this model
was the fact that it implements the vector boson mass
in a gauge invariant manner,
while the Higgs degrees of freedom with their thermal
masses of the order $\sim gT$ may be viewed as decoupling heavy particles
compared to the scale of interest $\sim g^2T$.
The maximal rate achievable in this model was found to be
$c\sim {\cal O}(0.01)$ for $m_0=0.1$,
but smaller by many orders of magnitude for other values of $m_0$.

Non-perturbative effects in finite temperature field theory
are related to the
infrared behaviour of the corresponding three-dimensional theory.
In a recent paper \cite{bp94} dynamical mass generation was studied
in a gauge invariant manner in
the three-dimensional SU(2) Higgs model.
Supplementing mass resummations with vertex resummations led
to a set of gauge independent gap equations for
the Higgs boson and vector boson masses. For
parameter values generally associated with the ``symmetric" phase, the
gap equations exhibit solutions with non-vanishing vacuum expectation
value and vector boson mass. This suggests that the ``symmetric" phase is
again a Higgs phase, just with modified parameters.
In subsequent work \cite{bp95} a comparison of the gap equation
approach as applied to the abelian and non-abelian Higgs model
established that the appearance of a non-vanishing vacuum expectation
value and vector boson mass in the ``symmetric" phase is specific to
the non-abelian model and hence not an artifact of the resummation
scheme.

In this letter I shall assume that the picture
developed in Ref.~\cite{bp94} gives a reliable description of the
``symmetric" phase and explore
its consequences for the sphaleron rate.
Given the structure of the resummed action,
the calculation of the rate in the modified loop expansion
is straightforward. Since the resummed three-dimensional theory
appears again to be in a Higgs phase, all that needs to be done is to match its
parameters properly to previous calculations of the rate in the
broken phase \cite{mcl90}-\cite{dia95}.
In this paper I work with the effective three-dimensional
theory without fermions, following the calculation in
Ref.~\cite{mcl90}.
This is particularly convenient because in this case the results of
\cite{bp94} can be used directly.
As will be discussed after the presentation of
this limiting case, a complete treatment including fermions
and keeping the non-zero Matsubara modes, analogous to that in
Ref.~\cite{dia95}, should also be possible in principle.

In order to illustrate the reduction of the rate calculation
in the ``symmetric" phase to the
known ones in the broken phase
it is necessary to
briefly outline the approach of Ref.~\cite{bp94}.
It is well known that for high temperatures and small enough
couplings
the SU(2) Higgs model may be
perturbatively reduced to yield an effective three-dimensional model
\beq \label{s3}
S_3 = \int d^3x \; \left[{1\over 4}F^a_{i j}F^a_{i j} +
(D_{i}\PH)^\hc D_{i}\Phi + \mu^2_3 \Phi^\hc \Phi
+  \lambda_3 (\Phi^\hc \Phi)^2 \right] \, ,
\eeq
where the effective parameters are related to the original
four-dimensional ones by \cite{bfh,fkrs}
\bea \label{match}
g_3^2 &=& g^2(T)T\ ,\,
\lambda_3 = \left(\lambda(T) - {3\over 128\pi}\sqrt{{6\over 5}}
g^3(T) + {\cal O}(g^4, \lambda^2)\right) T\ ,\nn\\
\mu^2_3 &=& \left({3\over 16} g^2(T) + {1\over 2}\lambda(T)
-{3\over 16\pi}\sqrt{{5\over 6}}g^3(T) + {\cal O}(g^4,
\lambda^2) \right) (T^2 - T_b^2) \, .
\eea
Here $T_b$ denotes the ``barrier temperature",
\beq
T_b=\left[\frac{16\lambda v^2}{3g^2(T)+8\lambda(T)}\right]^{1/2}\, ,
\eeq
which coincides with the critical temperature $T_c$ for a second-order
phase transition but is slightly lower than $T_c$ for a first-order
phase transition \cite{bb93}. Since the main interest here is in
temperatures well above the phase transition, this difference is
immaterial in what follows.
At tree level, for $\mu_3^2<0$, the Higgs field in
eq.~(\ref{s3}) develops a vacuum
expectation value $v_3^2=-\mu^2_3/\la_3$, the system is in the Higgs
phase, has a sphaleron solution and the loop expansion
in $g_3^2/m$ with $m=g_3v_3/2$ is convergent.
For $\mu_3^2>0$, $v_3=0$ and a loop expansion in terms of
the parameter $g_3^2/m$ is not possible.
This situation can be remedied by rearranging the perturbation series,
\beq
S_3=S_3+\delta S_3 - \delta S_3\equiv S_{3R}-\delta S_3\, ,
\eeq
where $\delta S_3$ contains mass and vertex
corrections from higher orders in the ordinary loop expansion.
In the modified expansion one-loop calculations are performed starting
from $S_{3R}$ while the terms included in $-\de S_3$ are treated as
counterterms which have to be taken into account in higher orders.
Calculating self-energies
in this scheme leads to a coupled set of non-linear
gap equations in which the physical pole masses of the vector boson ($m$)
and the Higgs boson ($M$) as well as the
the vacuum expectation value
of the Higgs field are determined self-consistently as functions of the
parameters of the theory,
\bea \label{rmass}
m^2&=&f(m/M,v_3;\;\la_3/g_3^2,\mu^2_3/g_3^4)\nn\\
M^2&=&g(m/M,v_3;\;\la_3/g_3^2,\mu^2_3/g_3^4)\nn\\
v_3&=&h(m/M,v_3;\;\la_3/g_3^2,\mu^2_3/g_3^4)\, ,
\eea
where the explicit forms of the functions $f,g,h$ are given in
Ref.~\cite{bp94}.
The resummed action may be written in terms of the solutions of these
equations as
\beq \label{lresum}
S_{3R} = \int d^3x\,\left[{1\over 4}F^a_{ij}F^a_{ij}
+ (D_i\Phi)^\hc D_i\Phi +\mu_{3R}^2 \Phi^{\hc}\Phi
+ \la_{3R} (\Phi^\hc\Phi)^2 \right] \, ,\\
\eeq
with the resummed parameters
\beq \label{rpar}
\la_{3R}=\frac{g^2M^2}{8m^2},\quad \mu^2_{3R}=-\frac{1}{2}M^2\, .
\eeq
{}From this form of the action it follows that the ``symmetric" phase
is again a Higgs phase, since $M^2$ is a physical mass and
always positive. The first-order phase transition with a jump in the
masses found for small values of $\la_3/g_3^2$ changes to a smooth crossover
at large $\la_3/g_3^2$.
This reorganisation of the perturbative series assumes that the
classical vacuum solution of the field equations
is not the true ground state of the theory, as is also indicated
by the breakdown of the loop expansion.
Instead, it perturbs around the true
vacuum containing quantum interactions.
Correspondingly, the classical field equations have no sphaleron
solution for $\mu_3^2>0$, while the field equations derived from
eq.~(\ref{lresum}) with $\mu^2_{3R}<0$ do have a non-trivial solution.
Clearly the form of these field equations and their sphaleron solution
is the same as that of the
classical ones in the broken phase, with the replaced parameters
$\la_3\rightarrow\lambda_{3R}$ and $\mu_3^2\rightarrow\mu^2_{3R}$.
In particular this means that the calculation of the rate now is the
same as that in the broken phase.

In order to transform
the action (\ref{lresum}) to the form used in
Ref. \cite{mcl90} coordinates and fields have to be
rescaled to be dimensionless,
\beq
x_i\rightarrow \xi_i/ 2m\,,\quad W_i\rightarrow 2 m W_i\, ,\quad
\Phi\rightarrow 2m \Phi\; .
\eeq
Using $m=g_3v_{3R}/2$ with $v_{3R}^2=-\mu^2_{3R}/\la_{3R}$ then leads to
\beq \label{rmcl}
S_{3R} =\frac{2m^2}{g_3^2} \int d^3\xi\,\left[{1\over 4}F^a_{ij}F^a_{ij}
+ (D_i\Phi)^\hc D_i\Phi
+ \frac{\la_{3R}}{g_3^2}\left(\Phi^\hc\Phi-\frac{1}{2}\right)^2 \right] \; .\\
\eeq
This action is now expanded in small fluctuations around a background
field,
\beq
W_i=W^b_{i}+w_i\, ,\quad \Phi=\Phi^b+\eta\, ,
\eeq
where the upper index $b$ denotes the vacuum or sphaleron solution and
$w_i,\eta$ are the fluctuation fields.
To eliminate gauge degrees of freedom, a gauge has to be fixed.
This is achieved by imposing the $R_{\xi=1}$ background gauge condition
\beq
D_i(W^b)w_i+\i(\Phi^\hc\tau\eta-\eta^\hc\tau\Phi)=0\; ,
\eeq
with the Fadeev-Popov determinant
\beq
\Delta_{FP}=\det\left[-D^2_i+\frac{1}{2}\Phi^\hc\Phi\right]\, ,
\eeq
and adding the corresponding gauge fixing and Fadeev-Popov terms to the
action,
\beq
S_{eff}=S_{3R}+S_{gf}+S_{FP}\, .
\eeq
In the Gaussian approximation only terms quadratic in the fields are
retained and
functional integration over the fields takes eq.~(\ref{langer})
to
\beq \label{rate1}
\frac{\Gamma}{V}=\frac{\mid \omega_-\mid}{2\pi}\frac{8\pi^2 N^3_{tr}N^3_{rot}}
{(g^2\beta)^3}{\rm e}^{-\beta E_S}(2 m \beta)^6 \kappa\, ,
\eeq
where $\kappa$ contains the fluctuation determinants
\beq
\kappa={\rm Im} \left(\frac{\det(\de^2S_{eff}/\de\phi^2)
\mid_{\phi=\phi_0}}
{\det'(\de^2S_{eff}/\de\phi^2)\mid_{\phi=\phi_S}}\right)^{1/2}
\left(\frac{\Delta_{FP}\mid_{\phi=\phi_S}}{\Delta_{FP}
\mid_{\phi=\phi_0}}
\right)\, .
\eeq
In this expression $\phi$ generically stands for the scalar and vector
fields of the theory, while the indices $S,0$ label sphaleron and vacuum
configurations, respectively. The prime on the determinant indicates
that the sphaleron translational and rotational
zero modes are to be omitted from its
evaluation. These modes have to be integrated separately by the method
of collective coordinates leading to the factor
$8\pi^2N^3_{tr}N^3_{rot}/(g^2\beta)^3$ in eq.~(\ref{rate1}), where
$N^3_{tr}$ and $N^3_{rot}$ are normalization integrals of the
zero modes depending on $\la_3/g_3^2$
(for a more detailed explanation
of the various factors see \cite{mcl90} and references therein).
Note that in this formula the ratio of Fadeev-Popov determinants enters
with a power of one as opposed to the calculation in \cite{mcl90},
where it enters with
a power of $1/2$. The reason is that in Ref.~\cite{mcl90} there are still
$W_0$'s in the theory. Gaussian integration over those fields
produces a factor $\Delta_{FP}^{-1/2}$.
In the current approach the $W_0$'s have been integrated out already in the
construction of the effective three-dimensional theory (\ref{s3})
and their contributions are included in its parameters.

The evaluation of eq.~(\ref{rate1}) is now straightforward.
The normalization factor $N^3_{tr}N^3_{rot}$, the negative mode $\om_-^2$,
the coefficient of the sphaleron energy $B$ and
the fluctuation determinants
$\kappa$ are all given in Ref.~\cite{mcl90} as functions
of $\la_3/g_3^2$. In the resummed scheme employed here, they are
functions of $\la_{3R}/g_3^2$ instead, depending on the original parameters
of interest
through the resummed ones according to eqs.~(\ref{rpar}), (\ref{rmass})
and (\ref{match}).
For any given set of parameters
$\{\la/g^2,T/T_b\}$ the matching equations (\ref{match}) give the
corresponding three-dimensional parameters $\{\la_3/g_3^2,\mu_3^2/g_3^4\}$.
These determine the solutions of the gap equations
which then lead to the required resummed parameters
$\{\la_{3R}/g^2_3,\mu^2_{3R}/g_3^4\}$
of the action (\ref{lresum}). The result of this procedure
is displayed in Fig.~1, where the temperature dependence of $\la_{3R}/g^2_3$
is shown for $\la/g^2=1/8$. Here and in the following
the slowly running gauge coupling was taken
to be a constant $g=0.67$ for simplicity.
The coefficient of the vector boson mass assumes the
constant value $m_0=0.28$ for $T>1.5 T_b$,
where it is also rather insensitive to variations
of $\la/g^2$ \cite{bp94}.
For any value of $\la_{3R}/g_3^2$ one may finally read off
all the quantities entering eq.~(\ref{rate1}) from the graphs given in
Ref.~\cite{mcl90}.
There are so far three complete calculations of the fluctuation
determinant $\kappa$.
The first of these \cite{mcl90}
seems to disagree with the other ones \cite{baa94,dia95}
for small values of $\la_3/g_3^2$. However, the results
of Ref.~\cite{baa94} and Ref.~\cite{dia95} agree within 10\% accuracy.
I have therefore chosen to take the data for the contributions to
$\kappa$ from Ref.~\cite{dia95}\footnote{I am grateful to
K.~Goeke and P.~Sieber for generously
providing me with their data}.

Denoting $\mid\om_-\mid/m$ by $\mid \bar{\om}_-\mid$,
the rate (\ref{rate1}) may be rewritten in the form of
eq.~(\ref{rate}) with
\beq \label{resc}
c=\mid\bar{\om}_-\mid 2^6(4\pi)^5 N^3_{tr}N^3_{rot}\, m_0^7\, \kappa\,
\exp[-8\pi m_0 B]\;.
\eeq
Note that, through its dependence on $\la_{3R}/g_3^2$ and the
matching equations,
$c$ is temperature dependent in general.
Fig.~2 shows the result of a numerical interpolation between several
data points obtained in the manner described above.
The solid line is the result for $\la/g^2=1/8$ or a zero temperature
tree-level mass ratio of $m^2_H/m^2_W=1$, and the dashed curve is for
$\la/g^2=1/64$ or $m^2_H/m^2_W=1/8$.
Close to the phase transition $c$
is strongly temperature dependent, dropping exponentially as $T_b$ is
approached.
For temperatures higher than $\sim
3T_b$ the sphaleron rate reaches its asymptotic form (\ref{rate})
with constant c, where it is also rather insensitive to the value
of the Higgs mass.
In this asymptotic region the constant coefficient takes on the
value $c\sim 0.01$. It should be emphasized that the onset of
the asymptotic
behaviour is a non-trivial prediction within the modified loop expansion,
relying on the temperature dependence of the resummed scalar coupling as
shown in Fig.~2.

Comparing with the results obtained earlier for the non-linear
sigma model \cite{op93} one finds that keeping the Higgs degree of
freedom enhances the rate considerably, even though it apparently has
little influence on the dynamics and the mass of the vector field
\cite{bp94}.
The reason is that the relevant saddle point solution of the
non-linear sigma model has, at the same value of $m_0$, substantially
higher energy than the
sphaleron solution of the Higgs model, resulting in a stronger
Boltzmann suppression. Therefore the Higgs degree of freedom may not
be neglected in a semiclassical treatment, even though it seems to play
only a marginal role dynamically.

In the case of $B-L=0$, the actual baryon number dissipation rate is
related to the sphaleron rate by \cite{amcl87}
\beq
\Gamma_B=\frac{dN_B}{N_Bdt}\approx
-\frac{13}{2}n_f\be^3\frac{\Gamma}{V}\; ,
\eeq
for $n_f$ fermion generations.
The temperature $T_*$ at which the sphaleron processes drop
out of equilibrium may be estimated by equating the dissipation rate
$\Gamma_B$
with the Hubble expansion rate of the universe.
Well above the phase transition the latter is $H\approx 16.6/(\be^2M_{Pl})$
\cite{ko90}
and one obtains
\beq
T_*\approx \frac{1}{16.6}\frac{13}{2}n_f M_{Pl}\alpha_W^4 c\sim
10^{11}{\rm GeV}\;.
\eeq

An important question concerns the accuracy of the numerical result
presented in Fig.~2.
Within the framework of the three-dimensional theory, there are
essentially two sources of possible corrections.
First, there might be higher order corrections to the masses calculated
from the resummed action, eq.~(\ref{rmass}).
The coefficient of the vector boson mass $m_0$ enters the prefactor
exponentially so that corrections to this
quantitiy are expected to alter the value of $c$ significantly.
Other changes, like small shifts in $\la_{3R}/g_3^2$ also due to
mass corrections, should only have a minor impact in comparison.
A quantitative estimate of the leading effect may be obtained by substituting
$m_0\rightarrow m_0(1+\gamma)$ in eq.~(\ref{resc}),
with $\gamma=\de m_0/m_0$ varying
between, say, $-0.5$ and $0.5$.
One finds that the prefactor
varies accordingly between $\sim 0.1$ and $\sim 0.0001$, respectively,
at a fixed temperature
$T=10T_b$. Or, to put it the other way round, the result
$c\sim {\cal O}(0.01)$ is stable as long as $\gamma\le 0.2$.
Hence, the order of magnitude of the coefficient $c$ cannot be considered
to be fixed
until the dynamical vector boson mass is determined with sufficient accuracy.
Second, there is the question of the applicability of the
Langer-Affleck formula (\ref{langer}).
If the saddle point expansion is to be valid,
the one-loop contribution should be small compared to the saddle point
contribution. In this calculation the fluctuations make up less than
$\sim 22\%$ of the saddle point energy, for all values of $T$.
This behaviour is improved for larger vector boson mass but worse
for smaller vector boson mass, which also follows from the
the effective loop expansion parameter, $g_3^2/m$.

Finally, the role of fermions and non-zero
Matsubara modes needs to be discussed.
The three-dimensional effective action (\ref{s3}) approximates the full
four-dimensional one up to terms of the
order $\sim m(T)/T$ \cite{fkrs},
where $m(T)$ now is a generic temperature
dependent mass $m\sim g^n T$ of any particle of the original theory.
For fermions, scalars and the zero component of the gauge field $n=1$,
while for the spatial vector bosons $n=2$ in the
``symmetric" phase, so corrections are suppressed by powers of coupling
constants. However, $g\sim 2/3$ is not very small and the corrections
might be important numerically. Moreover, since these are corrections to
the three-dimensional action which, up to a normlization, coincides
with the sphaleron energy functional, their
effect on the rate may be quite substantial.
In Ref.~\cite{dia95} the sphaleron fluctuation determinants were calculated
for the four-dimensional theory at finite temperature in the broken
phase, including also a doublet of fermions. Indeed it was found
that the fermion determinant enhances the sphaleron energy by
about $30\%$, leading to an additional suppression of the rate.
Since the scenario of the ``symmetric" phase discussed in this paper
resembles the broken phase in several respects,
a similar effect may occur here too. An investigation of the full
four-dimensional theory should be possible
in principle, albeit technically complicated. It would
require a separation of the zero and non-zero Matsubara modes which
after the resummation of the zero mode sector could not be treated
on the same footing anymore. Instead, a separate diagonalization of the
corresponding fluctuation operators would be necessary.

In summary, the gap equation approach suggested in Ref.~\cite{bp94}
was applied to the problem of B+L violation in the ``symmetric" phase
of the Standard Model
and found to be a useful calculational method. It yields the expected
asymptotic behaviour of the rate as well as the temperature where this
behaviour sets in, and the numerical coefficient of the rate.
The value of the coefficient turns out to be sensitive to higher order
corrections to the dynamical vector boson mass.
Given this uncertainty the numerical result obtained at one-loop
level may be regarded as compatible with previous results from
lattice Monte-Carlo simulations \cite{amb90}.
In order to tighten the results obtained here the underlying picture of the
``symmetric" phase needs to be confirmed by some other non-perturbative
method. Once this is achieved, one can go on and attempt to generalize to
the four-dimensional case including fermions.

\noindent
{\bf Acknowledgements:} I would like to thank K.~Goeke and P.~Sieber for
supplying me with data, and  W.~Buchm\"uller as well as
S.~Sarkar for a critical reading of the manuscript.

\section*{Figure captions}
\noindent
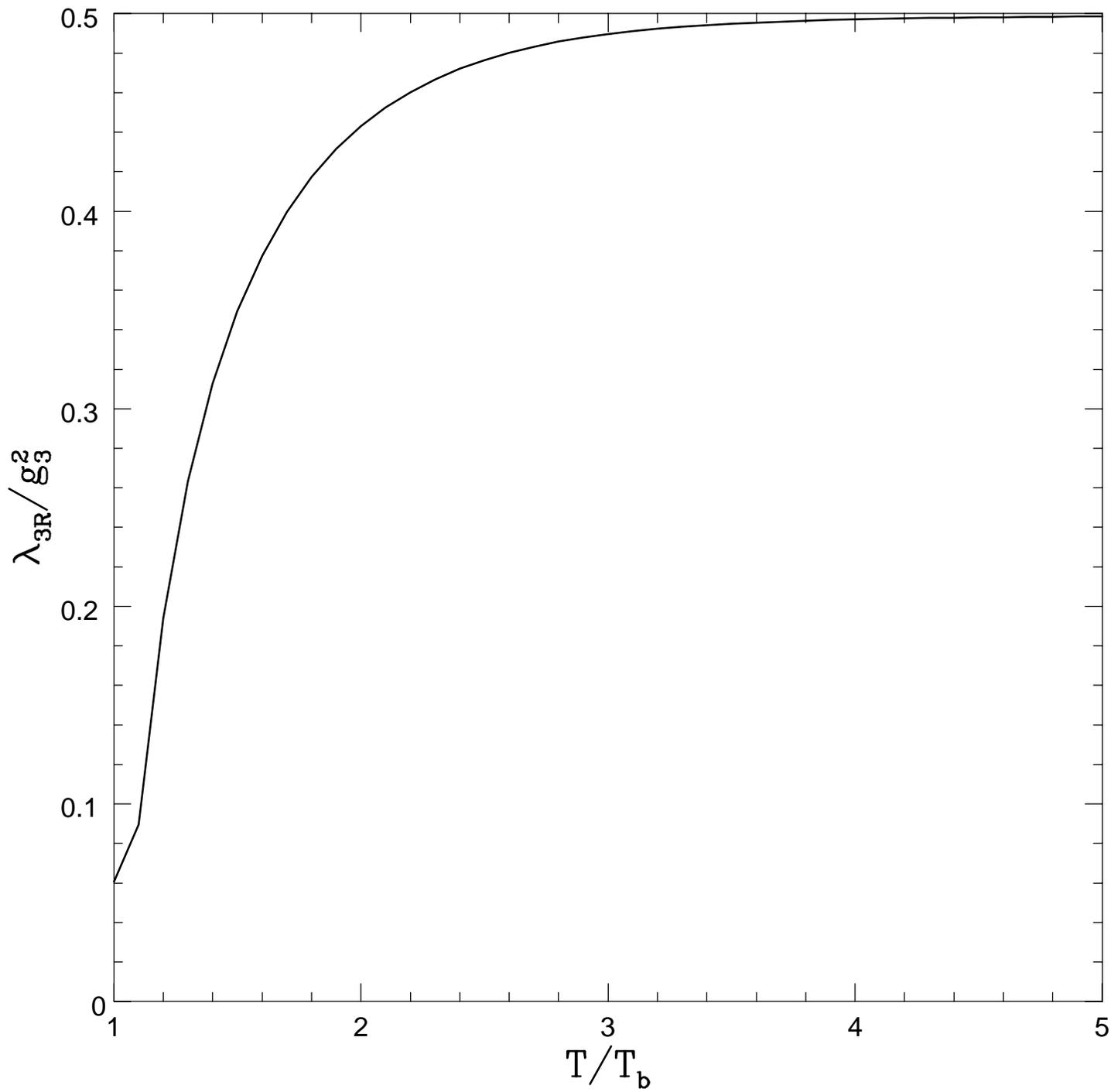
{\bf Fig.1} The resummed three-dimensional scalar coupling
$\la_{3R}/g^2_3$ as a function of temperature for $\la/g^2=1/8.$\\
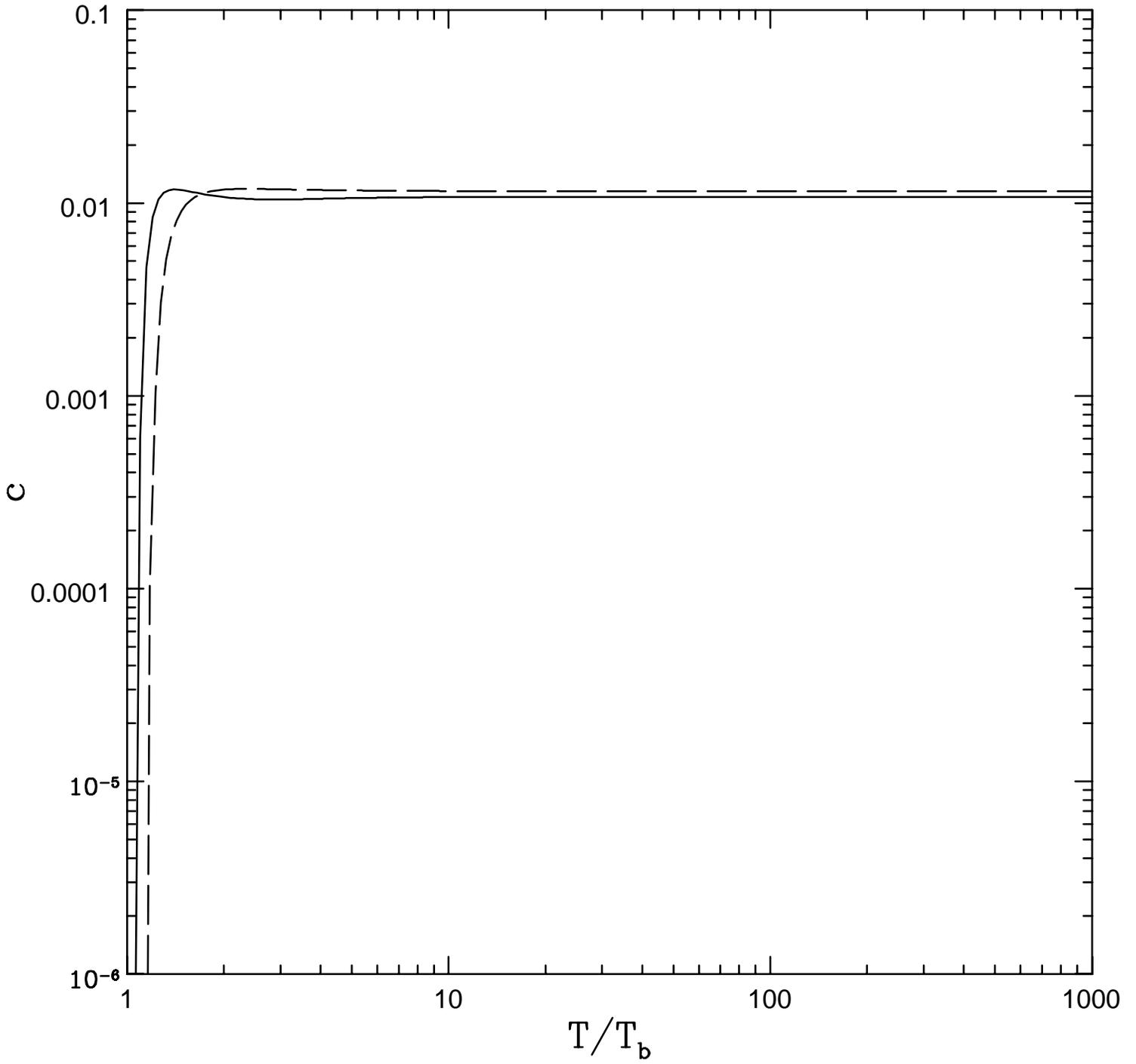
{\bf Fig.2} The coefficient $c$ of the sphaleron rate (\ref{rate})
as a function of temperature. The solid line corresponds to
$\la/g^2=1/8$ and the dashed line to $\la/g^2=1/64$.

\end{document}